\RequirePackage{fix-cm}
\documentclass[smallextended]{svjour3}       
\smartqed  
\usepackage{graphicx}
\usepackage{graphicx}
\usepackage{color, soul, xcolor}  
\usepackage{colortbl}
\usepackage{booktabs} 
\usepackage{enumitem}
\usepackage{caption}
\usepackage{amsmath}
\usepackage{listings}
\usepackage{balance}
\usepackage{caption}
\usepackage[caption=false,font=footnotesize,labelformat=simple]{subfig}
\usepackage[normalem]{ulem} 
\usepackage{framed}
\definecolor{shadecolor}{gray}{0.95}
\usepackage[most]{tcolorbox}
\usepackage[round]{natbib}   
\newcommand{\MYROMAN}[1]{%
  \textup{\uppercase\expandafter{\romannumeral#1}}%
}

\journalname{Empirical Software Engineering}

\begin{document}

\title{An Exploratory Characterization of Bugs in COVID-19 Software Projects}   

\titlerunning{Characterizing COVID-19 Software Bugs}        

\author{Akond Rahman         \and
             Effat Farhana 
}


\institute{Akond Rahman \at
              Department of Computer Science, Tennessee Technological University, Cookeville, TN, USA \\
              \email{arahman@tntech.edu}           
           \and
           Effat Farhana \at
              Department of Computer Science, North Carolina State University, Raleigh, NC, USA  \\
              \email{efarhan@ncsu.edu}
}

\date{\color{red}{The paper is under review.}}

\maketitle

\begin{abstract}

\textit{\ul{Context:}} The dire consequences of the COVID-19 pandemic has influenced development of COVID-19 software i.e., software used for analysis and mitigation of COVID-19. Bugs in COVID-19 software can be consequential, as COVID-19 software projects can impact public health policy and user data privacy. \textit{\ul{Objective:}} \textit{The goal of this paper is to help practitioners and researchers improve the quality of COVID-19 software through an empirical study of open source software projects related to COVID-19.}   
\textit{\ul{Methodology:}} We use 129 open source COVID-19 software projects hosted on GitHub to conduct our empirical study. Next, we apply qualitative analysis on 550 bug reports from the collected projects to identify bug categories.    
\textit{\ul{Findings:}} We identify 8 bug categories, which include data bugs i.e., bugs that occur during mining and storage of COVID-19 data. The identified bug categories appear for 7 categories of software projects including (i) projects that use statistical modeling to perform predictions related to COVID-19, and (ii) medical equipment software that are used to design and implement medical equipment, such as ventilators.  
\textit{\ul{Conclusion:}} Based on our findings, we advocate for robust statistical model construction through better synergies between data science practitioners and public health experts. Existence of security bugs in user tracking software necessitates development of tools that will detect data privacy violations and security weaknesses.


\end{abstract}


\section{Introduction}
\label{intro}

The novel Coronavirus disease (COVID-19) is a worldwide pandemic that spreads through droplets generated from coughs or sneezes and by touching contaminated surfaces~\citep{jhu:covid19:main}. As of May 31 2020, COVID-19 has caused 370,247 deaths across the world~\citep{jhu:covid19:main}. Apart from causing thousands of deaths and creating long term health repercussions for vulnerable populations, COVID-19 has severely impacted the economic sector. According to a recent study~\citep{stats:economy:covid19}, due to COVID-19 gross domestic product (GDP) will decrease from 3.0\% to 2.4\% worldwide. As of May 28 2020, nearly 41 million citizens reported unemployment in USA alone~\citep{stats:unemployment:covid19}. More than 3.9 billion people around the world were under some form of stay at home order due to COVID-19~\citep{news:lockdown}.

Health care professionals are at the frontline of combating COVID-19. Practitioners from other domains, such as software engineering have also joined forces to analyze and mitigate the negative consequences of COVID-19. For example, statistical modeling was used to build a software that identifies pneumonia caused by COVID-19 from lung scan images~\citep{china:ml:lungs:covid19}. The software was used in 34 Chinese hospitals~\citep{china:ml:lungs:covid19}. In response to the food insecurity caused by COVID-19, practitioners have created an interactive visualization software that displays free meal sites across USA~\citep{covid19:hunger:software}. The creators of the software envision in building a social movement to eradicate hunger and address economic inequalities. As another example, Apple and Google have jointly announced of creating a software framework that will help practitioners build tools to trace COVID-19 infection status of mobile app users~\citep{apple:tracing:api}. The above-mentioned examples show COVID-19 software i.e., software used for analysis and mitigation of COVID-19, to have near-term and long-term effects on public health and society.

\begin{figure}[t]
\centering
\includegraphics[scale=0.38]{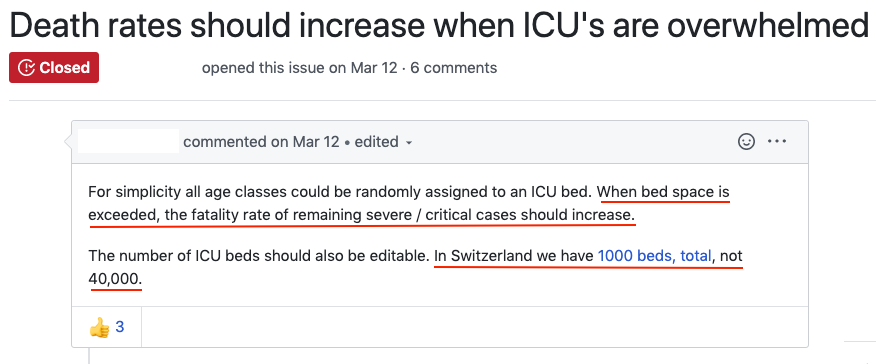} 
\caption{An example of a bug report related to statistical modeling in a software project called `neherlab/covid19\_scenarios'.}
\label{fig-bug-intro}
\end{figure}

Despite the above-mentioned advancements, COVID-19 software projects are susceptible to bugs. Let us consider Figure~\ref{fig-bug-intro} in this regard. Figure~\ref{fig-bug-intro} provides a snapshot of a bug report related to statistical modeling~\citep{bug:data:model}. We observe when implementing a statistical model the practitioners did not consider the correlation between ICU bed availability and death rate prediction. Furthermore, the number of intensive care unit (ICU) beds is incorrectly assumed to be 40,000 instead of 1,000. 

We hypothesize systematic analysis can reveal bug categories including statistical modeling bugs similar to Figure~\ref{fig-bug-intro}. In prior work researchers~\citep{Garcia:2020:AutoVehi, Rahman:2020:ACID, Linares-Vasquez:MSR2017, CATOLINOJSS2019:BugTypes, Thung:Lo:ISSRE2012, Wan:MSR2017} have documented the importance of bug categorization. For example, for autonomous vehicle software Garcia et al.~\citep{Garcia:2020:AutoVehi} stated that categorization of bugs can help to construct bug detection and testing tools. Linraes-V\'{a}squez et al.~\citep{Linares-Vasquez:MSR2017} stated categorizing vulnerabilities can help Android practitioners ``\textit{in focusing their verification and validation activities}''. According to Catolino et al.~\citep{CATOLINOJSS2019:BugTypes}, ``\textit{understanding the bug type represents the first and most time-consuming step to perform in the process of bug triage}''. Categorization of bugs in COVID-19 software can help practitioners and researchers to (i) understand the nature of COVID-19 software bugs, (ii) construct bug detection and repair tools, and (iii) measure COVID-19 software quality by using reported frequency of bug categories as a benchmark.

In prior work, researchers have categorized bugs for IaC~\citep{Rahman:2020:ACID}, autonomous vehicle~\citep{Garcia:2020:AutoVehi}, machine learning~\citep{Thung:Lo:ISSRE2012, Johir:FSE2019:Deep}, and blockchain~\citep{Wan:MSR2017} software. However, COVID-19 software is different from previously studied software in the following aspects: (i) \ul{\textit{development context}}: unlike previously studied software projects, COVID-19 software is developed in response to a pandemic that infected 6.1 million individuals in five months~\citep{jhu:covid19:main}, and (ii) \ul{\textit{public health}}: unlike previously studied software projects, COVID-19 software has direct implications on public health and relevant policy making for inhabitants in 188 countries.   

In response to the pandemic, researchers have conducted studies related to modeling~\citep{change:point:intro, yang:wang:ml, tamm:moscow}, biological science~\citep{jin:biology:intro, wang:antibody:intro, antivirus:intro, intro:neurologic}, social science~\citep{bavel:intro:social, tweet:intro:social, sports:intro:social, illness:intro:social, poland:intro:social}, and policy making~\citep{fauci:policy:intro, mello:policy:intro, rourke:intro:policy, mobility:policy:china}. However, characterization of bugs in COVID-19 software remains an unexplored area.

\textit{The goal of this paper is to help practitioners and researchers improve the quality of COVID-19 software through an empirical study of open source software projects related to COVID-19.}   

We answer the following research questions: 

\begin{itemize}
\item{\textbf{\textit{RQ1}: What categories of open source COVID-19 software projects exist?} 
}
\item{\textbf{\textit{RQ2}: What categories of bugs exist in open source COVID-19 software projects? How frequently do the identified bug categories appear? What is the resolution time for the identified bug categories?}  
}
\end{itemize}

\begin{figure}[h]
\centering
\includegraphics[scale=0.55]{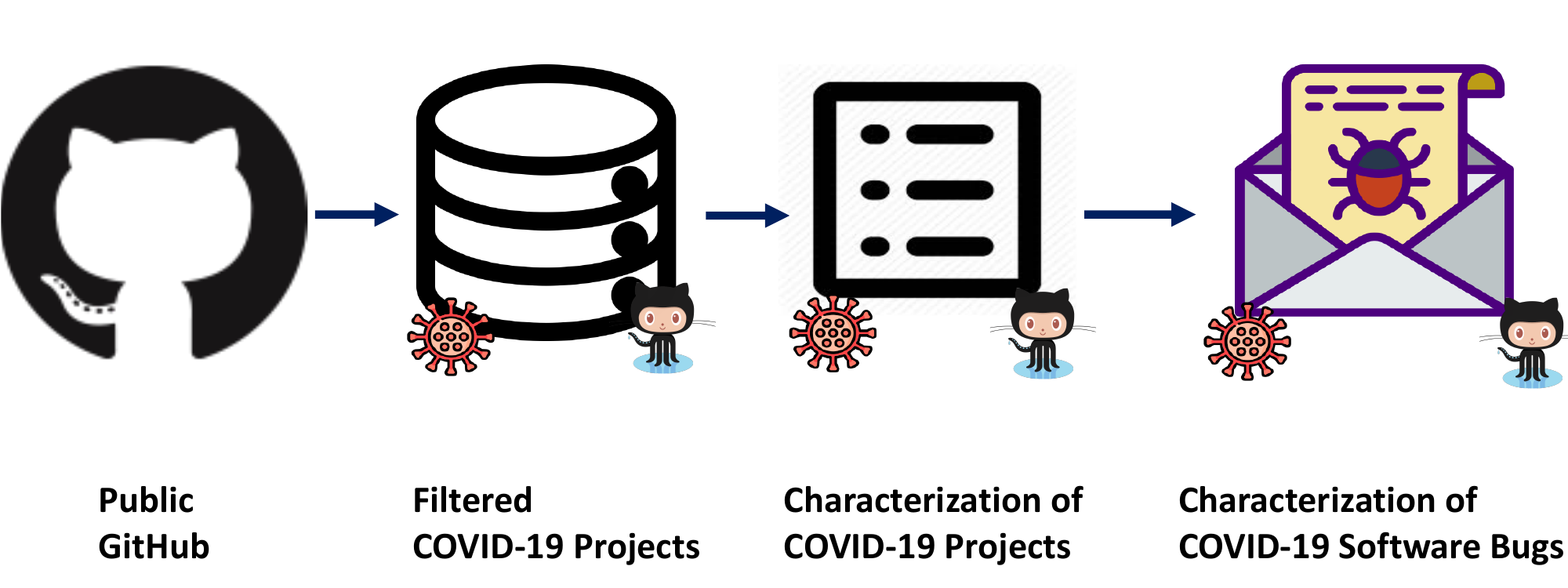}
\caption{An overview of our empirical study. }
\label{fig-meth-summary}
\end{figure}

We conduct an empirical study with 129 open source COVID-19 software projects hosted on GitHub. First, we apply qualitative analysis~\citep{saldana2015coding} on the README files of the collected open source software (OSS) projects to identify what categories of OSS projects exist related to COVID-19. Next, we apply qualitative analysis on 550 bug reports from the collected OSS projects to identify bug categories. We also quantify the frequency and resolution time of each bug category across the identified project categories. An overview of our paper is available in Figure~\ref{fig-meth-summary}.

\textbf{\textit{\ul{Contributions}}}: We list our contributions as following: 

\begin{itemize}
\item{A categorization of bugs that appear in COVID-19 software projects}; 
\item{A categorization of OSS projects related to COVID-19}; 
\item{An empirical study that identifies what category of bugs appear for what category of COVID-19 software projects}; and 
\item{A curated dataset which maps each identified bug report to the identified bug categories~\footnote{https://figshare.com/s/7044678e1d7e7feb1efb} }. 
\end{itemize}

We organize rest of the paper as following: we provide background and discuss related work in Section~\ref{bg-rel}. We provide the methodology and results for RQ1 and RQ2 respectively, in Sections~\ref{rq1} and~\ref{rq2}. We discuss our results with a summary of our findings in Section~\ref{discussion}. We provide the limitations of our paper in Section~\ref{threats}. Finally, we conclude the paper in Section~\ref{conclusion}. 

\section{Background and Related Work}
\label{bg-rel}

In this section, we first provide background on COVID-19 in Section~\ref{bg} and briefly describe related research in Section~\ref{rel}.  

\subsection{Background}
\label{bg}

COVID-19 stands for `Coronavirus disease 2019'~\citep{jhu:covid19:main}. COVID-19 is an infectious disease that causes severe respiratory problems for infected human beings. The first case of an infected COVID-19 patients was reported in December 2019 in Wuhan, China~\citep{jhu:covid19:main}. Since then the disease has spread rapidly. To date 6,108,525 cases have been reported across 188 countries, which resulted in 370,247 deaths~\citep{jhu:covid19:main}.  

COVID-19 is highly contagious~\citep{covid19:contagious} and is caused by a strain of Coronavirus called severe acute respiratory syndrome Coronavirus 2 (SARS-CoV-2)~\citep{covid19:sars-cov-2}. Disagreements exist amongst researchers on how the SARS-CoV-2 was transmitted to human beings. By using genome sequence similarity, a group of researchers speculate that the virus was transmitted to human beings from pangolins~\citep{wong:tx:pangolin}. Researchers have found the SARS-CoV-2 virus to have 92\% resemblance with the Coronavirus found in pangolins~\citep{zhang:pangolin:covid19}. Pangolins are nocturnal mammals found in Asia and Sub-Saharan Africa. Other researchers speculate the SARS-CoV-2 to be transmitted from horseshoe bats and civets. Researchers from McMaster University found the SARS-CoV-2 virus to share 99.8\% of it's genome with a civet Coronavirus~\citep{nature:covid19:civet}. From another phylogenetic analysis, researchers observed that a virus from horseshoe bats have a 96\% resemblance to the SARS-CoV-2 virus~\citep{nature:covid19:bat}.

COVID-19 is a human transmissible disease. Transmission occurs via respiratory droplets from coughs and sneezes within a range of $\mathtt{\sim}$6 feet. The virus can also be transmitted to a human being via contaminated surfaces as well as through droplets generated from cough and sneezing. What makes COVID-19 more susceptible for transmission is that infected human beings do not exhibit symptoms up until 2$\mathtt{\sim}$14 days from infection~\citep{jhu:covid19:main}. Symptoms of COVID-19 include cough, shortness of breath, fever, sore throat, and loss of taste or smell.   

As of May 31, 2020, no vaccines exist that can cure COVID-19. To prevent the spreading, health experts have strongly recommended for maintaining personal hygiene e.g., frequent hand washing, social distancing, and mandatory lockdown, when necessary~\citep{jhu:covid19:main}. 

Researchers~\citep{harvard:recurrence:2024}~\citep{china:recurrence} have provided evidence that support the recurring nature of COVID-19. About the recurrence of COVID-19 Kissler et al.~\citep{harvard:recurrence:2024} stated ``\textit{a resurgence in contagion could be possible as late as 2024.}''.        

\subsection{Related Work}
\label{rel}

Our paper is related with prior research that has focused on computing research related COVID-19 and characterization of bugs in OSS projects. We briefly describe each of these areas in the following subsections: 

\subsubsection{Computing Research Related to COVID-19}
\label{rel-covid19}

Our paper is related to recent computing research on COVID-19. Since the outbreak of COVID-19 in December 2019, researchers have conducted extensive research on understanding the spread of the disease through lens of computer science. We describe related work as following: 

\begin{itemize}[leftmargin=*] 
\item{\textit{Visualization techniques}: Dey et al.~\citep{dey:viz} constructed a visualization tool called Visual Exploratory Data Analysis (V‐EDA) to understand the spread of COVID-19.   
}
\item{\textit{Machine learning and statistical modeling}: Naude~\citep{naude:ai} identified research areas where machine learning can be applied to combat COVID-19: (i) early notifications, (ii) tracking and prediction, (iii) data dashboards, (iv) diagnosis and prognosis, (v) treatments and cures, and (vi) social control. Researchers~\citep{yang:wang:ml, tamm:moscow} in separate studies have used statistical models to understand the COVID-19 outbreak. Yang and Wang~\citep{yang:wang:ml} proposed a mathematical model to understand the COVID-19 outbreak in Wuhan, where the virus originated from. They observed the diseases to be endemic and advocated for long-term disease prevention and intervention public health programs. Tamm~\citep{tamm:moscow} constructed a mathematical model to understand the outbreak in Moscow using five scenarios based on control measures. Tamm~\citep{tamm:moscow} reported that fatality due to COVID-19 would remain extremely high and healthcare providing institutions would be overloaded. Randhawa et al.~\citep{randhawa:ml} applied machine learning to classify pathogens, and observed evidence that supports the hypothesis that COVID-19 originated from a bat, as their model classified the COVID-19 virus as `Sarbecovirus', a sub-category within `Betacoronavirus'. Rao and Vasquez~\citep{rao:vazquez:ml} used machine learning algorithms to identify potential COVID-19 patients using mobile web-based survey data. Currie et al.~\citep{currie:ml} identified challenges in modeling the COVID-19 pandemic, which included quarantine strategies and case isolation, social distancing measures and applications, lock down management, and testing for COVID-19. Pandey et al.~\citep{Pandey:ML:tracing} used machine learning to construct a contract tracing mobile application for COVID-19 that uses the smartphone's senor data and self-assessment of the smartphone user. Santosh~\citep{santosh:active:learning} advocated for the usage of active learning and multi-modal data to identify COVID-19 outbreaks as the pandemic is world-wide and differences in location can impact model performance used in forecasting.   
}
\item{\textit{Robotics}: Yang et al.~\citep{yang:robots} advocated for the usage of robotics to help combating COVID-19 as robots can be deployed to deliver food and medicine, as well as for disinfecting infrastructure, such as medical centers and schools.  
}

\end{itemize} 

The above-mentioned discussion shows a lack of research that have characterized bugs in COVID-19 software projects. We address this research gap in our paper. 

\subsubsection{Research Related to Bug Characterization} 
\label{rel-swe}

Our paper is also related with prior research that have characterized bugs in OSS. Mockus et al.~\citep{mockus:herbshleb:tosem:mozilla} studied the contribution nature in OSS Apache and Mozilla projects. They~\citep{mockus:herbshleb:tosem:mozilla} observed contributors who submit bug reports are approximately 8.2 times higher in number than contributors who address bugs in bug reports. Ma et al.~\citep{ma:crossproject:bugs} investigated Python GitHub projects that are used in the scientific domain, and observed developers to use stack traces, as well as communicate with upstream developers, to identify root causes of bugs. Zhang et al.~\citep{bug:report:github:diffs} examined bug reports for mobile and desktop software hosted on GitHub, and identified differences on how the reports are constructed. Ray et al.~\citep{Ray:FSE2014:Lang} studied the correlations between bugs and the language the software is being developed, and reported a modest correlation using an empirical study of 729 GitHub projects. Categorization of domain-specific OSS bugs has also been investigated: Thung et al.~\citep{Thung:Lo:ISSRE2012}, Garcia et al.~\citep{Garcia:2020:AutoVehi}, Wan et al.~\citep{Wan:MSR2017}, Islam et al.~\citep{Johir:FSE2019:Deep}, and Rahman et al.~\citep{Rahman:2020:ACID} in separate research papers used OSS projects to classify bug categories respectively, for machine learning, autonomous vehicle, blockchain, deep learning, and IaC.    

We take motivation from above-mentioned research and study COVID-19 software bugs in the following manner: 

\begin{itemize}  
\item{categories of bugs;}
\item{frequency of identified bug categories;}
\item{resolution time of identified bug categories; and}
\item{categories of software projects}. 
\end{itemize}

\section{RQ1: COVID-19 Software Project Categories}
\label{rq1}

In this section, we answer ``\textit{\textbf{RQ1: What categories of open source COVID-19 software projects exist?}}''. We define COVID-19 software projects as software projects used for analysis and mitigation of COVID-19. We hypothesize multiple categories of COVID-19 software projects to exist in the OSS domain. We validate our hypothesis by systematically categorizing COVID-19 software projects. Our categorization will provide insights on how the software development community has responded to the COVID-19 pandemic.


\subsection{Methodology for RQ1}
\label{meth-rq1}

We answer RQ1 by completing the following steps: 

\subsubsection{Dataset Collection}
\label{meth-rq1-repo-collection} 

We conduct our empirical analysis by collecting COVID-19 software projects hosted on GitHub. To collect these projects we use GitHub's search utility~\citep{github:covid19:search}, where we first identified projects tagged as `covid-19'.  We use the search string `covid-19', as it is a topic designated for COVID-19 by GitHub~\citep{github:covid19:topic}. Upon collection of the projects we apply a set of filtering criteria so that we can identify projects that contain sufficient data for analysis. We describe the filtering criteria below: 

\begin{itemize}[leftmargin=*] 

\item{

\textbf{Criterion-1}: The project must have at least 2 developers. Our assumption is that this criterion will filter projects used for personal purposes.   

}

\item{

\textbf{Criterion-2}: The project has at least 5 open issues. We use this filtering criterion to identify projects that are actively maintained.  

}

\item{

\textbf{Criterion-3}: The project must have at least two commits per month. Munaiah et al.~\citep{MunaiahCuration2017} used the threshold of at least two commits per month to determine which projects have enough development activity for software organizations. We use this threshold to filter projects with short development activity.

}

\item{

\textbf{Criterion-4}: The README of the project is written in English. README projects related to COVID-19 can be non-English. We do not include non-English projects as raters who will perform categorization are not familiar with non-English languages, such as Spanish and Cantonese. 

}

\item{

\textbf{Criterion-5}: The project is actually related to COVID-19. Practitioners can mislabel projects using the `topic' feature of GitHub. For example, from manual inspection we observe the `RehanSaeed/Schema.NET'~\footnote{https://github.com/RehanSaeed/Schema.NET} project to be tagged as `covid-19', even though it is used to convert blob objects into C\# classes.

}

\end{itemize} 

\subsubsection{Qualitative Analysis of README files}
\label{meth-rq1-qual}

We apply a qualitative analysis called open coding~\citep{saldana2015coding} on the content of README files for each of the downloaded projects from Section~\ref{meth-rq1-repo-collection}. README files describe the content of the project and give GitHub users an overview of the software project~\citep{treude:readme}. We hypothesize that by systematically analyzing the content of the README files we can derive what types of software projects are developed that are related to COVID-19.  

\begin{figure}[]
\centering
\includegraphics[scale=0.57]{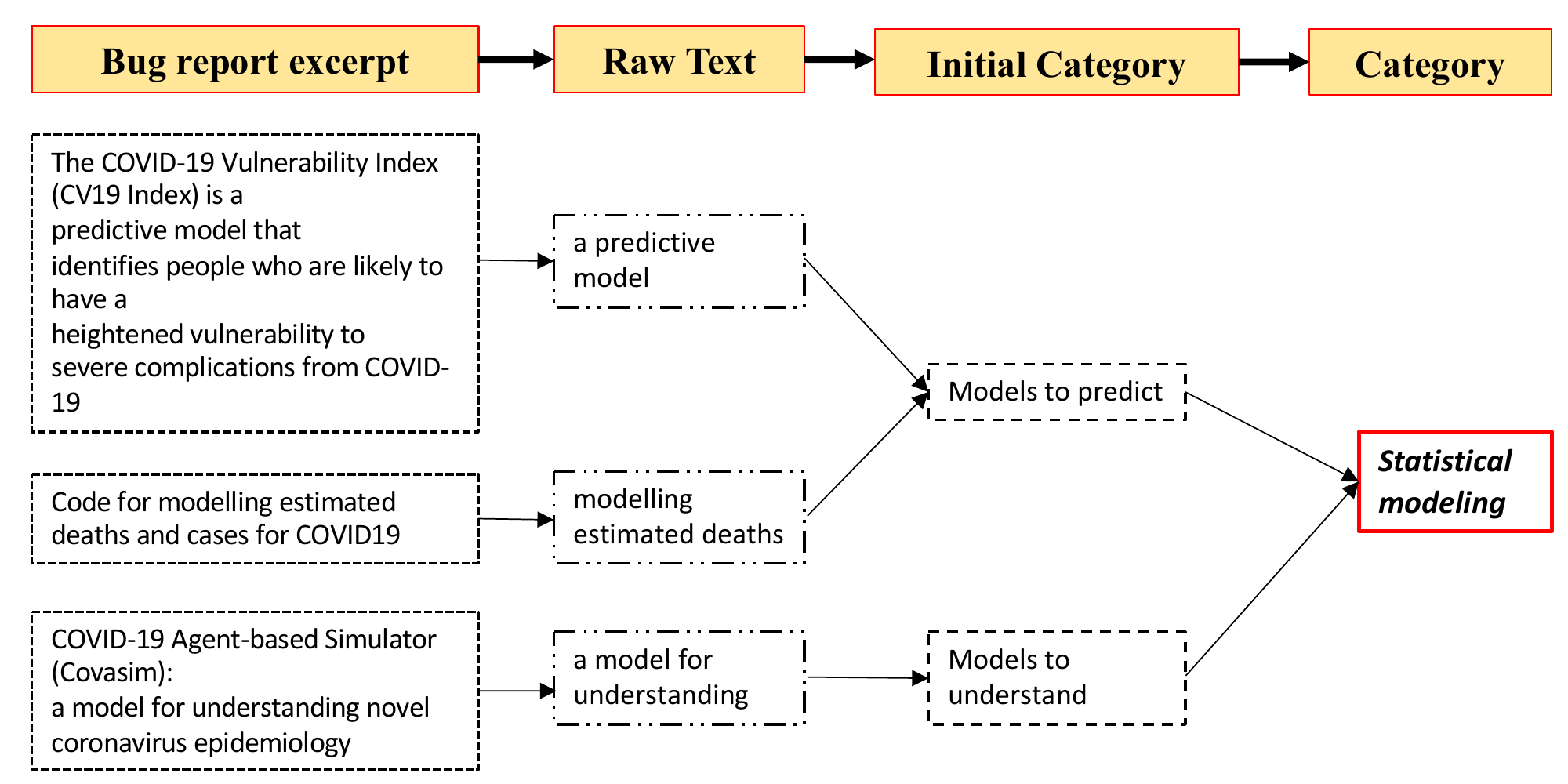}
\caption{A hypothetical example to demonstrate our process of open coding to categorize COVID-19 software projects.} 
\label{meth-rq1-open-coding}
\end{figure}

In open coding a rater identified and synthesizes patterns within unstructured text~\citep{saldana2015coding}. We select open coding because we can obtain detailed information on the software project categories. We use a hypothetical example to demonstrate our process of open coding in Figure~\ref{meth-rq1-open-coding}. First, we collect text from the README files for each of the collected projects from Section~\ref{meth-rq1-repo-collection}. Next, we extract text snippets that describe the purpose of the software project. For example, from the raw text `\textit{The COVID-19 Vulnerability Index (CV19 Index) is a predictive model that identifies people who are likely to have a heightened vulnerability to severe complications from COVID-19}' we extract the text snippet `\textit{a predictive model}', as the extracted text snippet describes the purpose of the software project. Next, from the text snippets `\textit{a predictive model}' and `\textit{modelling estimated deaths}' we generate an initial category called `\textit{Models to predict}'. Two initial categories `\textit{Models to predict}' and `\textit{Models to understand}' are combined into one category `\textit{Statistical modeling}', as they both indicate the descriptions of the software projects to be related with statistical modeling. 

The first and second author conduct the open coding process separately. The first and second author respectively an experience of 10 and 6 years in software engineering and has experience in conducting open coding upon software project artifacts, such as commit messages~\citep{Rahman:2020:ACID} and Stack Overflow posts~\citep{effat:icsme2019}. Upon completion of open coding process the first and second author identify agreements and disagreements. Disagreements are resolved upon discussion, Agreement rate is calculated using Cohen's Kappa~\citep{cohens:kappa}. During the discussion phase both authors agreed present their justification, and recheck the category derivation based on the discussion and revisiting content. The mapping determined upon discussion is considered final. One project can map to multiple categories.

\subsubsection{Closed Coding}
\label{meth-rq1-closed coding}

We apply closed coding~\citep{crabtree:coding:book} to identify which project maps to the identified categories from Section~\ref{meth-rq1-qual}. Closed coding is the qualitative analysis techniques where a rater maps an artifact to a pre-defined category by inspecting the artifact~\citep{crabtree:coding:book}. The first and second author separately conduct closed coding on the collected README files. After completing the closed coding process the first and second authors identify agreements and disagreements. Agreement rate is recorded using Cohen's Kappa~\citep{cohens:kappa}. Disagreements are resolved using discussion. During the discussion phase both authors present their justification for disagreements. Next, based on the discussion the authors recheck the labeling based on the justification and content analysis. The categorization determined upon discussion is considered final.    

\subsubsection{Rater Verification}
\label{meth-rq1-rater-verification} 

The derived categories are susceptible to the bias of the first and second author. We mitigate the limitation by allocating an additional rater who applied closed coding for a subset of the README files. The additional rater who is not an author of the paper, is a fourth year PhD student in the Department of Computer Science at Tennessee Technological University. The rater has a professional experience of 2 years in software engineering and has conduced qualitative analysis on software artifacts, such as bug reports. We randomly allocate a set of 100 README files mined from 100 projects to the rater. The rater applies closed coding on the content of the README files, to identify the mapping between each project and identified categories. Upon completion of closed coding we calculate Cohen's Kappa~\citep{cohens:kappa} between the rater and the first author, as well as with the second author, separately.   


\subsection{Answer to RQ1}
\label{res-rq1}

We answer RQ1 by first providing summary statistics of our dataset in Section~\ref{res-rq1-dataset}. Next, we report categories of the projects in Section~\ref{res-rq1-categs}.

\subsubsection{Summary of Dataset}
\label{res-rq1-dataset} 

Altogether we download 129 projects for analysis. Using the search feature we identify 3,276 public projects upon which we apply our filtering criterion. A complete breakdown of our filtering criterion is available in Table~\ref{table-rq1-repos-filtering}. Summary statistics of the projects is available in Table~\ref{table-rq1-repos-attr}. `Languages' in Table~\ref{table-rq1-repos-attr} correspond to the count of main programming languages of the collected projects as determined by GitHub's linguist tool~\citep{github:linguist}. Example languages include JavaScript, Python and R.  

\begin{figure}[h]
\centering
\includegraphics[scale=0.75]{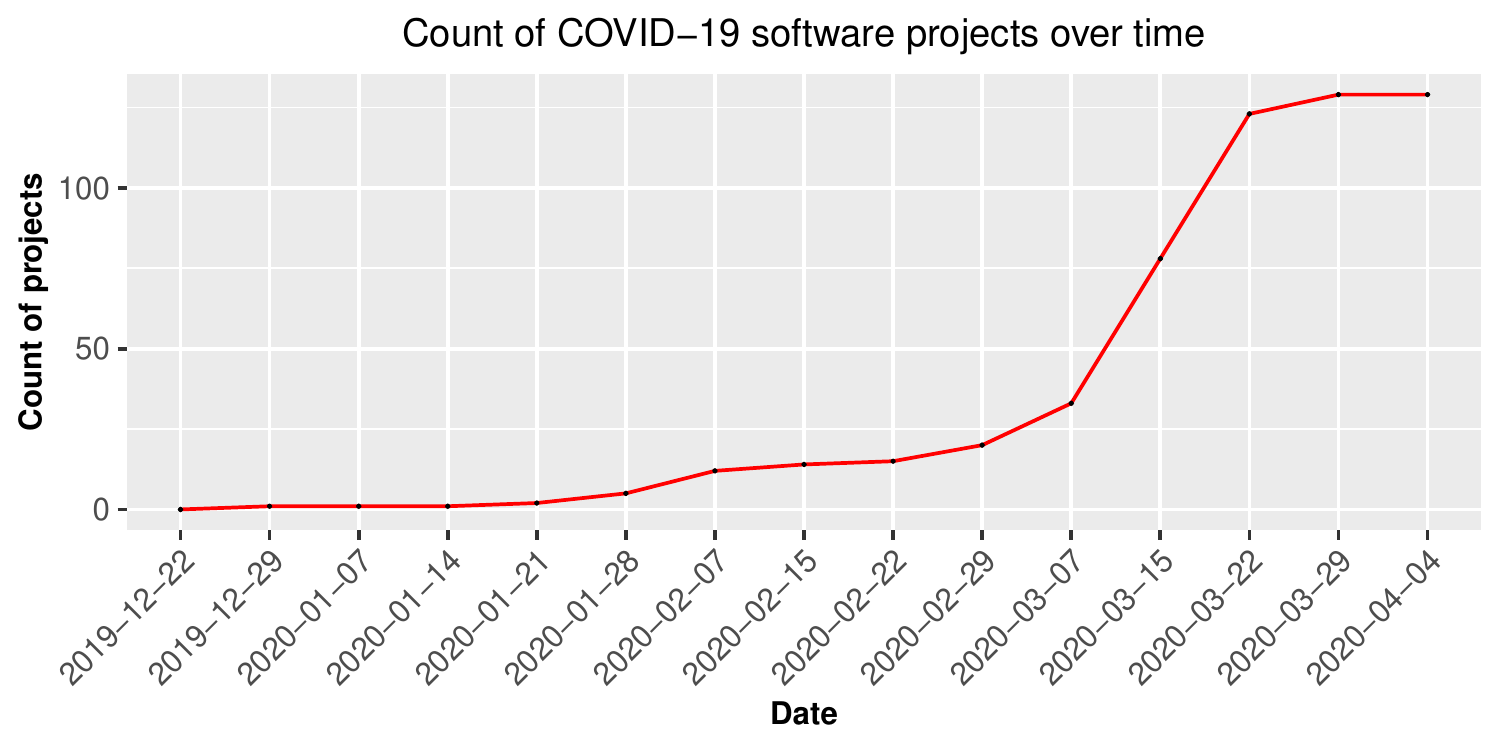}  
\caption{Temporal evolution of COVID-19 software projects based on their creation date. We observe sharp increase in project creation after Feb 29, 2020.} 
\label{fig-rq1-temporal}
\end{figure}

A temporal evolution of the 129 COVID-19 software projects based on creation date is available in Figure~\ref{fig-rq1-temporal}. We observe sharp increase in project creation after Feb 29, 2020.     

\begin{table}[]
\centering
\caption{Filtering of COVID-19 projects used in paper.}
\label{table-rq1-repos-filtering}
{\footnotesize
\begin{tabular}{ p{7.5cm}  r  }
\hline
\textbf{Criteria}                                                  & GitHub  \\
\hline
\textbf{Initial}                                                     &  3,276  \\
\hline
\textbf{Criterion-1 (Devs $>= 2$)}                   &  1,287  \\
\textbf{Criterion-2 (Open issues $>= 5$)}      &  169  \\
\textbf{Criterion-3 (Commits/month $>=2$)} &  154  \\
\textbf{Criterion-4 (README is English)}        &  131  \\
\textbf{Criterion-5 (Actually COVID-19)}        &  129  \\
\hline
\textbf{Final}                                                       & 129  \\
\hline
\end{tabular}
}
\end{table}

\begin{table}[]
\centering
\caption{Attributes of studies COVID-19 projects.}
\label{table-rq1-repos-attr}
{\footnotesize
\begin{tabular}{ p{3.5cm}  r }
\hline
\textbf{Attributes}     & Total \\
\hline
\textbf{Commits}        & 38,152  \\
\textbf{Developers}     & 2,243  \\
\textbf{Duration}       & 12/2019-03/2020  \\
\textbf{Files}          & 24,839  \\
\textbf{Issues}         & 4,405   \\
\textbf{Languages}      & 18  \\
\textbf{Releases}       & 286  \\
\textbf{Projects}       & 129  \\
\hline
\end{tabular}
}
\end{table}  

\subsubsection{Categorization of COVID-19 Software Projects} 
\label{res-rq1-categs} 

We identify 7 categories of COVID-19 software projects. We describe each of the categories below in an alphabetic order:  

\includegraphics[width=0.3cm, height=0.25cm]{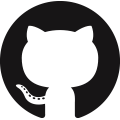} \textbf{\MYROMAN{1}}: \textbf{\textit{\ul{Aggregation}}}:: This category includes software projects that curate data related to COVID-19 and present collected COVID-19 data in an aggregated format using visualizations. The purpose of these projects is to help users understand the spread of the COVID-19 disease over time and location. Software projects that belong to this category can be country specific as done in `juanmnl/covid19-monitor'~\citep{repo:juanmnl} and `dsfsi/covid19za'~\citep{repo:covid19za} respectively, for Ecuador and South Africa. Aggregation of COVID-19 data can also be at a global level, for example, `boogheta/coronavirus-countries'~\citep{repo:boogheta} is a software that aggregates COVID-19 data across the world and allows software users to compare the reported cases on a country-by-country basis.    

\includegraphics[width=0.3cm, height=0.25cm]{gh.png} \textbf{\MYROMAN{2}}:  \textbf{\textit{\ul{Education}}}:: This category includes projects that provide utilities on educating people about COVID-19. Lack of knowledge related to infections and symptoms can contribute to rapid spreading of COVID-19. The purpose of these projects is to build software, where users can ask questions and obtain answers. We observe two categories of software: \textit{first}, question and answer websites similar to Stack Overflow~\footnote{https://stackoverflow.com/}, such as `nthopinion/covid19'~\citep{repo:nthopinion}, where users can ask questions about COVID-19, and other users answer such questions. \textit{Second}, we observe bot-specific software, such as `deepset-ai/COVID-QA'~\citep{repo:deepset-ai} that provides answers for questions related to COVID-19 automatically.  

\includegraphics[width=0.3cm, height=0.25cm]{gh.png} \textbf{\MYROMAN{3}}: \textbf{\textit{\ul{Medical equipment}}}:: This category includes projects to curate and maintain source code for the design and implementation of medical equipment used to treat COVID-19. The purpose of these projects is to create designs of COVID-19 related medical equipment, such as ventilators at scale, so that the growing need of medical equipment in hospitals is satisfied. One example of such repository is `makers-for-life/makair'~\citep{repo:makers-for-life}, which states the following in it's README page: ``\textit{Aims at helping hospitals cope with a possible shortage of professional ventilators during the outbreak. Worldwide. ... We target a per-unit cost well under 500 EUR, which could easily be shrunk down to 200 EUR or even 100 EUR per ventilator given proper economies of scale, as well as choices of cheaper on-the-shelf components}''.  The project includes design of the proposed ventilators as CAD files, as well as relevant firmware available as C++ code files. 

Another example is the `popsolutions/openventilator'~\citep{repo:popsolutions}, which aims to provide cheap but reliable ventilators to treat COVID-19 in economically under-developed regions of the world. The software project initiated from a Facebook group called `Open Source COVID-19 Medical Supplies'~\footnote{https://www.facebook.com/groups/opensourcecovid19medicalsupplies/}, where members discussed the scarcity of ventilators and the importance of creating cheap ventilators through efficient design. In the project we notice developers to create, build, and share designs using OpenSCAD scripts. OpenSCAD is an open source tool to build computer-aided design (CAD) objects~\footnote{https://www.openscad.org/}.    

\includegraphics[width=0.3cm, height=0.25cm]{gh.png} \textbf{\MYROMAN{4}}: \textbf{\textit{\ul{Mining}}}:: This category includes projects that provide APIs to mine COVID-19 data from data sources, such as the US Center for Disease Control and Prevention (CDC)~\citep{data:cdc}, the World Health Organization (WHO)~\citep{data:who}, and data reported from local institutions. The purpose of this category of software is to provide utilities for software developers so that they can get real-time access to COVID-19 data to build aggregation software, discussed above. Because of the nature of the pandemic, access to real-time data is  pivotal for accurate aggregation and analysis. The mining tools help developers to get such support. Mining software can be location specific, for example `dsfsi/covid19africa'~\citep{data:dsfsi} is dedicated to curate and collate COVID-19 related data for African countries.  

\includegraphics[width=0.3cm, height=0.25cm]{gh.png} \textbf{\MYROMAN{5}}: \textbf{\textit{\ul{User tracking}}}:: This category includes software projects that collects information from users regarding their COVID-19 infection status. Tracking of user information can happen voluntarily, where the user voluntarily self reports COVID-19 infection status. The `enigmampc/SafeTrace'~\citep{contract:tracing:covid:enigmampc} software is an example where users self report their infection status as well as location history. Tracking of user information can also be done using inference, as done in `OpenMined/covid-alert'~\citep{contract:tracing:covid:alert}, where the software collects user's location information to predict if the user is in a location with high infection density. One utility of these projects is to identify high-risk locations so that users can have an understanding of which nearby location can be avoided. Self reporting software have yielded benefits for China and South Korea~\citep{contract:tracing:east:asia}.

\includegraphics[width=0.3cm, height=0.25cm]{gh.png} \textbf{\MYROMAN{6}}:  \textbf{\textit{\ul{Statistical modeling}}}:: This category includes software that use statistical models to predict attributes related to COVID-19. The purpose of the projects is to make predictions for the future based on existing data. Example usage of statistical models include (i) predicting death rate as done in `ImperialCollegeLondon/covid19model'~\citep{repo:type:model:ImperialCollegeLondon}, (ii) automating the process of lung segmentation with computerized tomography (CT) scan, as done in `JoHof/lungmask'~\citep{repo:type:model:JoHof}, (iii) predicting the impact of the COVID-19 pandemic on hospital demands as done in `neherlab/covid19\_scenarios'~\citep{repo:type:model:neherlab}, and (iv) predicting presence of COVID-19 with X-ray images using deep learning as done in `elcronos/COVID-19'~\citep{repo:type:model:elcronos}.   

\includegraphics[width=0.3cm, height=0.25cm]{gh.png} \textbf{\MYROMAN{7}}: \textbf{\textit{\ul{Volunteer management}}}:: This category includes software used to efficiently manage volunteering effort. The purpose of this software is to build software platforms so that users can volunteer and participate in activities to help distressed families and communities. One example is the `covid-volunteers'~\citep{repo:type:vol:helpwithcovid} software, which provides a web portal where users can sign up for 650 projects that include donation of masks, personal protective equipment (PPEs), and testing of COVID-19~\footnote{https://helpwithcovid.com/medical}. Platforms can be global, such as `covid-volunteers', and also regional, for example `Applifting/pomuzeme.si'~\citep{repo:type:vol:mgmt} creates a web portal so that people inside Czech Republic can volunteer.   

\subsubsection{Frequency of the Identified Categories} 
\label{res-rq1-categs-freq} 

Based on project count aggregation is the most frequent category. Along with project count, we provide summary statistics of projects that belong to each category in Table~\ref{tab-res-rq1-categ-atts}. We also observe on average user tracking projects to be more frequently released compared to other project types. 

We identify four software projects that belong to multiple categories. As an example, the `soroushchehresa/awesome-coronavirus'~\citep{repo:soroushchehresa} project belongs to the categories: aggregation, mining, and statistical modeling. 

\begin{table*}[]
\captionsetup{justification=centering}
\caption{Summary statistics of projects that belong to each category. Based on project count `Aggregation' is the most frequent category. }  
\label{tab-res-rq1-categ-atts}  
{\footnotesize
\begin{tabular}{p{1.75cm} p{1.25cm}  p{1.25cm} p{1.25cm}  p{1.25cm}  p{1.1cm}   p{1.1cm}  }
\hline
\textbf{Proj. Categ.}  & \textbf{Projects}\ & \textbf{Com.}  & \textbf{Devs}     & \textbf{Files}      & \textbf{Iss.} & \textbf{Rele.} \\  
\hline
         \textbf{Aggregation}   & \cellcolor{green!50} 50              & 14,985         & 663               & 8,641               & 908           & 72                    \\ 
         \textbf{Mining}        & 35              & 9,671          & 894               & 6,714               & 515           & 21                     \\ 
         \textbf{Stat. model.}  & 22              & 7,214          & 429               & 3,464               & 491           & 38                     \\ 
         \textbf{Education}     & 9               & 4,550          & 196               & 1,696               & 406           & 14                     \\ 
         \textbf{User track}    & 9               & 2,020          & 152               & 2,291               & 119           & 286                   \\ 
         \textbf{Volunteer.}    & 7               & 2,186          & 143               & 2,041               & 320           & 0                      \\ 
         \textbf{Med. equip.}   & 3               & 859            & 38                & 790                 & 14            & 63                     \\ 

\hline          
\end{tabular}
}
\end{table*}

\subsubsection{Rater Agreement} 
\label{res-rq1-rater} 

We report agreement rate for three steps: open coding, closed coding, and rater verification. 

\noindent \textit{Open coding}: After completing open coding, the first and second author respectively, identified 7 and 10 categories. The agreement rate is 70.0\%, and the Cohen's Kappa is 0.7, indicating `substantial' agreement~\citep{Landis:Koch:Kappa:Range}. The authors disagreed on `Volunteering software related to local communities', `Education bots', and `Aggregated visualizations', additional categories identified the second author. Upon discussion both authors agree that the category `Education bots' can be merged with `Education' as it encompasses all categories of knowledge software, such as bots and web applications. The authors also agreed that `Volunteering software related to local communities' can be merged with `Volunteer management' and `Aggregated visualizations' can be merged with `Aggregation', as `Aggregation' includes software that aggregates COVID-19 data and displays aggregated data with visualizations.

\noindent \textit{Closed coding}: During closed coding the first and second author mapped each of the 129 project to an existing category. The agreement rate is 93.8\%. The Cohen's Kappa is 0.92. The authors disagreed on the labeling of 8 projects, which are resolved through discussion.

\noindent \textit{Rater verification}: We also measured the agreement rate between an additional rater and the authors for categorizing README files of projects. Cohen's Kappa between the additional rater and the first author for a randomly selected set of 50 README files is 0.73, indicating `substantial' agreement~\citep{Landis:Koch:Kappa:Range}. Cohen's Kappa between the additional rater and the second author for a randomly selected set of 50 README files is 0.73, indicating `substantial' agreement~\citep{Landis:Koch:Kappa:Range}. The agreement rate between the additional rater and the first and second author is respectively, 78.0\% and 76.0\%.

\section{RQ2: Bug Categories} 
\label{rq2}

In this section, we answer ``\textit{\textbf{RQ2: What categories of bugs appear in COVID-19 software projects? How frequently do the identified bug categories appear? What is the resolution time for each bug category?}}'' A categorization of bugs for COVID-19 software projects can inform practitioners and researchers about how software related to COVID-19 is developed and in which areas they can help. Furthermore, educators can learn about the software bugs that occur in a software related to a pandemic and disseminate these findings in the classroom. Frequency of the identified bug categories can help us understand what categories of software tend to contain what types of software bugs and provide quality improvement suggestions accordingly. Quantifying the resolution time for bugs in software projects can help software engineering researchers provide actionable guidelines to practitioners. For example, Wan et al.~\citep{Wan:MSR2017} observed that for blockchain software projects security bugs can take longer to fix compared to other bug categories. Based on their findings Wan et al.~\citep{Wan:MSR2017} recommended that blockchain project maintainers can adopt security analysis and repair tools to fix security bugs quickly.


\subsection{Methodology to Identify Bug Categories, Frequency, and Resolution Time} 
\label{meth-rq2}

In this section we provide the methodology to identify bug categories, quantify bug category frequency, and bug resolution time. 

\textbf{Methodology to Identify Bug Categories}\label{meth-rq2-categories}: We identify bug categories using the following steps: 

\begin{itemize} 
\item{\textit{Step\#1-Filtering}: We collect the 4,405 issue reports from the 129 projects and manually inspect each issue report. We do not rely on automated approach, such as keyword search or using bug labels, as automated approaches tend to generate false positives, which may bias research results~\citep{herzig:bug:feature:icse2013}. While inspecting each issue report we use the following IEEE definition for bugs: ``\textit{an imperfection that needs to be replaced or repaired}''~\citep{ieee:def}, similar to prior work~\citep{Rahman:2020:ACID}. By completing this step we will obtain a set of closed issues reports that correspond to bugs. We use closed reports because as open bug reports are often incomplete and may not help in identifying bugs~\citep{Wan:MSR2017}.      

The first and second author manually inspect individually to identify what issue reports correspond to bugs. We record agreement rate and Cohen's Kappa~\citep{cohens:kappa} between the first and second author. Disagreements between the first and second author is resolved through discussions. The process is subjective and susceptible to the bias of the first and second author. We mitigate the bias by using an additional rater, who inspected randomly inspected 100 issue reports and classified them as bug reports and non-bug reports. The additional rater is the fourth year PhD student at Tennessee Technological University who is also involved in rater verification for RQ1.   
}

\item{\textit{Step\#2-Open coding}: We apply open coding~\citep{saldana2015coding} on the content of the collected bug reports from \textit{Step\#1}. Our open coding process for deriving bug categories is similar to our process of deriving project categories illustrated in Figure~\ref{meth-rq1-open-coding}. First, we extract raw text from bug report titles and description, from which we generate initial categories. Next, we merge initial categories based on the commonalities and generate categories. 

Similar to deriving project categories, the first and second author separately apply the process of open coding to generate bug categories. Upon completion of the process we quantify agreement rate and measure Cohen's Kappa~\citep{cohens:kappa}. For disagreements we conduct discussion. Generated categories upon discussion is considered final. 

}

\end{itemize}

\textbf{Methodology to Quantify Bug Category Frequency}\label{meth-rq2-freq}: We apply the following steps to quantify the frequency of identified bug categories: 

\begin{itemize} 

\item{\textit{Step\#1-Closed coding}: We apply closed coding~\citep{crabtree:coding:book} to map each identified category to the bug reports that we study. The first and second author separately apply closed coding for the collected bugs from \textit{Step\#1}. Upon completion, we calculate the agreement rate and Cohen's Kappa~\citep{cohens:kappa}. Disagreements are resolved using discussion. 
     
}

\item{\textit{Step\#2-Metric calculation}: We quantify the frequency of the identified bug categories using two metrics: `Proportion of Bugs Across All Projects (BugPropAll)' and `Proportion of Bugs For a Certain Project Category (BugPropCateg)'. We use Equations~\ref{equ:bug:prop:all} and~\ref{equ:bug:prop:categ} to respectively calculate `BugPropAll' and `BugPropCateg'. The `BugPropAll' metric using Equation~\ref{equ:bug:prop:all} provides a holistic overview of the frequency of identified bug categories. The `BugPropCateg' metric using Equation~\ref{equ:bug:prop:categ} provides a granular overview of bug category frequency for each software project types identified from Section~\ref{res-rq1-categs}.    
}

\item{\textit{Step\#3-Rater verification}: The use of first and second author as raters to conduct closed coding is susceptible to rater bias. We mitigate this limitation by allocating an additional rater. We assign randomly selected 250 bug reports to the additional rater who apply closed coding. We provide the additional rater with a document that provides definitions of each identified category with examples. 

Similar to our process of rater verification for project categorization, the additional rater is the fourth year PhD student in the Department of Computer Science in Tennessee Tech. University. The fourth year PhD student is involved in the rater verification process for identifying project categories and labeling issue reports as bug reports.       
}

\end{itemize}

\begin{equation}\label{equ:bug:prop:all}
\begin{aligned}
\text{BugPropAll($x$)} = \\ \frac{\text{\# of bug reports labeled as category $x$}}{\text{total \# of bug reports}}*100\%
\end{aligned}
\end{equation} 

\begin{equation}\label{equ:bug:prop:categ}
\begin{aligned}
\text{BugPropCateg($x, y$)} = \\ \frac{\text{\# of bug reports labeled as $x$, belonging to project type $y$}}{\text{total \# of bug reports for project type $y$}}*100\%
\end{aligned}
\end{equation}

\textbf{Methodology to Quantify Bug Resolution Time} \label{meth-rq2-duration} We use the open and closing timestamp for each closed bug report in our dataset to quantify the resolution time for each bug category, similar to Wan et al.~\citep{Wan:MSR2017}. We calculate bug resolution time by computing the number of hours that have elapsed between when the bug report is opened and closed, and not re-opened again, as per our dataset , which was downloaded on April 04, 2020. We report bug resolution time for all bug categories, as well as for bug reports that belong to certain categories of software projects. 


\subsection{Answer to RQ2}
\label{res-rq2}

We answer RQ2 by first providing a breakdown of how we obtained our bug reports in Table~\ref{table-rq2-bugs-filtering} and~\ref{tab-res-rq2-repo-categ-bugs}. As shown in Table~\ref{tab-res-rq2-repo-categ-bugs}, the categories with the most and least bug reports is respectively, aggregation and medical equipment. One project can belong to multiple categories, and that is why the total count of bug reports do not total to 550. On average, bug reports per project to vary from 1.3$\mathtt{\sim}$6.4, as shown in the `BugReport/Project' column.    

\begin{table}[]
\centering
\caption{Filtering of bug reports from COVID-19 software projects.}
\label{table-rq2-bugs-filtering}
{\footnotesize
\begin{tabular}{ p{7.5cm}  r  }
\hline
\textbf{Initial}                            &  4,095  \\
\hline
\textbf{Criterion-1 (Closed issues)}        &  2,965  \\
\textbf{Criterion-2 (Valid bug reports)}    &  550  \\
\hline
\textbf{Final}                              & 550  \\
\hline
\end{tabular}
}
\end{table}

\begin{table}[]
\centering
\caption{Count of bug reports for each category of COVID-19 software projects. Aggregation-related projects have the highest amount of bug reports. }   
\label{tab-res-rq2-repo-categ-bugs}  
{\footnotesize
\begin{tabular}{p{4cm} r  r }
\hline
\textbf{Project category}           & \textbf{Count (\%)}              & \textbf{BugReport/Project}\\  
\hline
         \textbf{Aggregation}       & \cellcolor{green!50} 220 (40\%)  & 4.4\\ 
         \textbf{Mining}            & 150 (27.3\%)                     & 4.3\\ 
         \textbf{Stat. Model.}      & 98  (17.8\%)                     & 4.4\\ 
         \textbf{Education}         & 58  (10.5\%)                     & 6.4\\ 
         \textbf{Volunteer.}        & 40  (7.3\%)                      & 4.4\\ 
         \textbf{User Track}        & 31  (5.6\%)                      & 4.4\\ 
         \textbf{Med. Equip.}       &  4  (0.7\%)                      & 1.3\\ 

\hline          
\end{tabular}
}
\end{table}  

Next, we describe the identified bug categories in Section~\ref{res-rq2-categ} by applying open coding on the collected 550 bug reports. The frequency of the identified bug categories is provided in Section~\ref{res-rq2-freq}. We provide details of rater verification in Section~\ref{res-rq2-rater}. Finally, we provide the bug resolution time in Section~\ref{res-rq2-duration}.

\subsubsection{Bug Categories of COVID-19 Projects}  
\label{res-rq2-categ} 

We identify 8 bug categories, which we describe below alphabetically: 


\includegraphics[width=0.35cm, height=0.3cm]{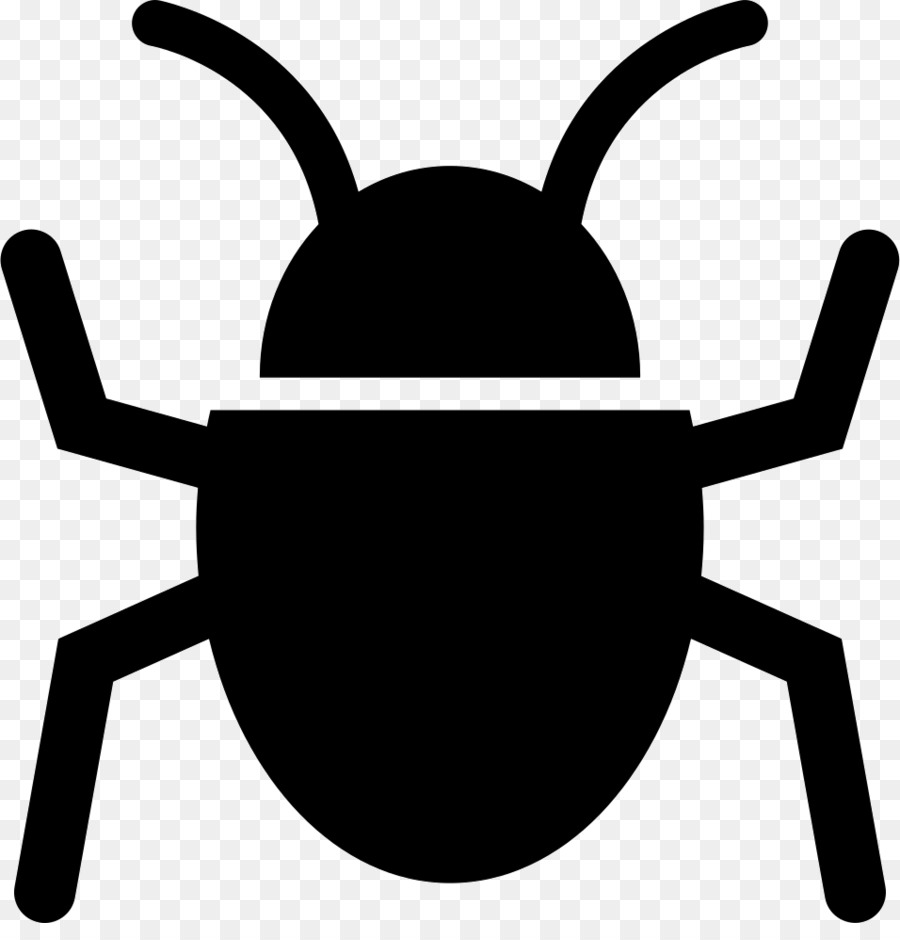} \textbf{\MYROMAN{1}}:  \textbf{\textit{\ul{Algorithm}}}:: This category corresponds to bugs when implementation of an algorithm does not follow expected behavior. An algorithm is a sequence of computational steps that transform input into output~\citep{cormen:algo:book}. We observe algorithm bugs to include two sub-categories: (i) bugs related to statistical modeling algorithms, where statistical modeling results are incorrect due to incorrect assumptions and/or implementations, and (ii) bugs related to incorrect logic implemented in the software. Algorithm bugs have been previously observed in autonomous vehicle software~\citep{Garcia:2020:AutoVehi} and machine learning software~\citep{Thung:Lo:ISSRE2012}.  

\textit{\textbf{Example}}: We provide examples for the two sub-categories: 

\begin{itemize}

\item{Statistical modeling: In a bug report titled ``\textit{Death rates should increase when ICU's are overwhelmed}''~\citep{bug:data:model}, a practitioner describes how incorrect assumption can result in incorrect modeling behavior. The practitioner discusses that bed space is correlated with estimation of fatality rate. When bed space of hospitals are exhausted hospitals will not be able to treat new COVID-19 new patients, which could potentially increase the fatality rate. 

The bug report provides evidence that if the context of COVID-19 is not correctly incorporated in statistical models, them models will provide incorrect results. Incorrect statistical models can be consequential, as countries are adopting public health policies specific to COVID-19. For example researchers have critiqued the statistical models derived by the Institute for Health Metrics and Evaluation at the University of Washington (IHME), and advised USA policy makers to use the modeling results with caution~\citep{bug:data:model:news}.     

}

\item{Incorrect logic: In a bug report titled ``\textit{Fix Prefecture Sorting}''~\citep{bug:data:algo}, a practitioner describes a sorting bug which occurs when trying to visualize COVID-19 cases based on prefectures in Japan. A prefecture is an administrative jurisdiction in a country similar to a state or province~\citep{prefecture}. The bug occurred due to an incorrect logic that did not perform sorting by prefectures. 

} 

\end{itemize}


\includegraphics[width=0.35cm, height=0.3cm]{bug.jpg} \textbf{\MYROMAN{2}}: \textbf{\textit{\ul{Data}}}:: This category corresponds to bugs that occur during mining and storage of COVID-19 data. As discussed in Section~\ref{res-rq1-categs} we observed our dataset to include projects that mine and aggregate COVID-19 data. We observe four sub-categories of data bugs: (i) \ul{storage}: bugs that occur while storing data in a database, (ii) \ul{mining}: bugs that occur while retrieving data from data APIs, (iii) \ul{location}: bugs where location information in stored data is incorrect, and (iv) \ul{time series}: bugs that correspond to missing data for a certain time period. Data bugs have been previously reported for deep learning software~\citep{Johir:FSE2019:Deep}.  

\textit{\textbf{Example}}: We provide examples for each of these sub-categories below:

\begin{itemize}

\item{Storage: In a bug report titled ``\textit{Temperature data not saved in the backend}''~\citep{bug:data:storage}, a practitioner describes a bug where patient temperature data is inserted in the front-end but not stored in the database. 

}

\item{Mining: When mining data from sources bugs occur. A practitioner describes a mining bug in a bug report titled ``\textit{CDC Children scraper is outdated}''~\citep{bug:data:mining}. The mining tool mines data related to children affected by COVID-19. 

}

\item{Location: in a bug report titled ``\textit{Rajasthan District names are wrong}'', a practitioner describes that inserted location data for an Indian state called `Rajasthan' is wrong~\citep{bug:data:location}.

}

\item{Time series: missing data was reported for a project and reported in a bug report titled ``\textit{Data has a gap between 2020-3-11 and 2020-3-24}''~\citep{bug:data:time}.  

}

\end{itemize}


\includegraphics[width=0.35cm, height=0.3cm]{bug.jpg} \textbf{\MYROMAN{3}}:  \textbf{\textit{\ul{Dependency}}}:: This category corresponds to bugs that occur when execution of the software is dependent on a software artifact that is either missing or incorrectly specified. For COVID-19 projects an artifact can be an API or a build artifact. Dependency bugs were previously reported in IaC~\citep{Rahman:2020:ACID}, machine learning~\citep{Thung:Lo:ISSRE2012}, and audio processing software~\citep{Cotroneo:TestType:JSS2013}.  

\textit{\textbf{Example}}:  In a bug report titled ``\textit{Missing PostGIS}''~\citep{bug:dep}, a practitioner describes that installation and execution of the software is prohibited due to a software package called `PostGIS', which is used to store spatial and geographic measurements, such as area, distance, polygon, and perimeter in PostgreSQL databases. 


\includegraphics[width=0.35cm, height=0.3cm]{bug.jpg} \textbf{\MYROMAN{4}}:  \textbf{\textit{\ul{Documentation}}}:: This category corresponds to bugs that occur when incorrect and/or incomplete information in specified in release notes, maintenance notes, and documentation files, such as README files. Documentation bugs have been reported in prior research related to autonomous vehicle~\citep{Garcia:2020:AutoVehi} and IaC~\citep{Rahman:2020:ACID}.  

\textit{\textbf{Example}}: In a bug report titled ``\textit{Missing code of conduct}'', a practitioner describes a `CODE\_OF\_CONDUCT.md' file to be missing in a Markdown file that describes how practitioners can contribute to the project~\citep{bug:doc}. 


\includegraphics[width=0.35cm, height=0.3cm]{bug.jpg} \textbf{\MYROMAN{5}}:  \textbf{\textit{\ul{Performance}}}:: This category corresponds to bugs that cause performance discrepancies for the software. Performance bugs are manifested in slow response of the web or mobile app. Performance bugs have been previously reported in prior research related to blockchain software~\citep{Wan:MSR2017}.  

\textit{\textbf{Example}}: In a bug report titled ``\textit{Cluster animation slowing down the browser. It also takes much time}'', a practitioner describes how a performance bug related to an animation feature is slowing down a Firefox browser on Windows 10~\citep{bug:perf}. The performance bug was reported for a website called `covid19india.org'~\footnote{https://www.covid19india.org/}, which aggregates COVID-19 data for India and displays them.   


\includegraphics[width=0.35cm, height=0.3cm]{bug.jpg} \textbf{\MYROMAN{6}}:  \textbf{\textit{\ul{Security}}}:: This category corresponds to bugs that violate confidentiality, integrity, or availability for the software. Prior research on bug categorization have also observed security bugs to appear for blockchain software~\citep{Wan:MSR2017}, video game software~\citep{penta:games:msr2018}, and IaC~\citep{Rahman:2020:ACID}.  

\textit{\textbf{Example}}: In a bug report titled ``\textit{Fix password reset procedure}''~\citep{bug:security}, a practitioner describes a password reset bug, where the password reset procedure ends arbitrarily after 500 login attempts.  


\includegraphics[width=0.35cm, height=0.3cm]{bug.jpg} \textbf{\MYROMAN{7}}:  \textbf{\textit{\ul{Syntax}}}:: This category corresponds to bugs related with the syntax of the programming languages used to develop the software. Syntax-related bugs have been reported in prior research related to IaC~\citep{Rahman:2020:ACID}, deep learning~\citep{Johir:FSE2019:Deep}, and OSS GitHub software~\citep{Ray:FSE2014:Lang}.  

\textit{\textbf{Example}}: We notice bugs related to data types in `neherlab/covid19\_scenarios'. In the bug report titled ``\textit{Fix types and linting errors}''~\citep{bug:syntax}, a practitioner describes how linting and type checking was disabled for the project, which led to bugs related to linting and type checking. 


\includegraphics[width=0.35cm, height=0.3cm]{bug.jpg} \textbf{\MYROMAN{8}}: \textbf{\textit{\ul{UI}}}:: This category corresponds to bugs that involve the user interface (UI) of the software. UI bugs include navigation-related bugs on web pages, bugs related to accessibility, displaying incorrect images, links, and color, and responsiveness. UI bugs have been previous reported for blockchain software~\citep{Wan:MSR2017}.  

\textit{\textbf{Example}}: In a bug report titled ``\textit{accessibility fixes}''~\citep{bug:ui:accessibility} describes a UI bug related to accessibility. According to the bug report, a screen reader incorrectly renders check marks and crosses in front of the ``\textit{Do's and Don't as M's and N's}''.

\subsubsection{Frequency of Identified Bug Categories} 
\label{res-rq2-freq} 

Based on the `Proportion of Bugs Across All Projects (BugPropAll)' metric we observe UI bugs to be the most frequent category, whereas documentation is the least frequent category. We provide a compete breakdown of the metric in Table~\ref{tab-rq2-freq-all}. Data bugs have four sub-categories: storage, mining, location, and time series. The frequency for storage, mining, location, and time series is respectively, 4.7\%, 5.8\%, 87.2\%, and 2.3\%. Algorithm bugs have two sub-categories: statistical modeling and wrong logic. The frequency for statistical modeling and wrong logic is respectively, 42.3\% and 57.7\%.

\begin{table}[]
\centering
\caption{Frequency of identified bug categories. UI-related bugs are the most frequent.}
\label{tab-rq2-freq-all}
{\footnotesize
\begin{tabular}{ p{4cm}  r }
\hline
\textbf{Bug category}   & \textbf{BugPropAll (\%)}  \\
\hline
\textbf{UI}             & \cellcolor{green!50} 38.2  \\
\textbf{Data}           & 30.9  \\
\textbf{Dependency}     & 18.9  \\
\textbf{Algorithm}      & 7.8 \\
\textbf{Syntax}         & 6.7  \\
\textbf{Security}       & 2.5  \\
\textbf{Performance}    & 1.6   \\
\textbf{Documentation}  & 1.4  \\
\hline
\end{tabular}
}
\end{table}  

We observe bug category frequency to vary for different categories of projects. We provide the `Proportion of Bugs For a Certain Project Category (BugPropCat)' values for each project category in Table~\ref{tab-rq2-freq-categ}. `AGG', `MINE', `STA', `EDU', `TRAK', `VOL' and `EQU' respectively, corresponds to the seven project categories: aggregation, mining, statistical modeling, education, user tracking, volunteer management system, and medical equipment. 

According to Table~\ref{tab-rq2-freq-categ}, except for mining and medical equipment software, the dominant bug category is UI. One possible explanation can be the analyzed software projects have UIs, which may have contributed to the frequency of UI bugs. For mining software the dominant bug category is data bugs i.e., bugs that occur due to storing and processing of COVID-19 data. For medical equipment software the dominant bug category is dependency. We also notice algorithm bugs to be the second most frequent bug category for statistical modeling software. Similar to prior work on machine learning~\citep{Thung:Lo:ISSRE2012}, we expected algorithm bugs to be the most dominant category for statistical modeling. Statistical modeling software also have UIs for user interaction, and the count of UI bugs may have foreshadowed the count of algorithm bugs.

\begin{table*}[]
\captionsetup{justification=centering}
\caption{Bug category frequency for each identified project type. All values are presented in (\%). }  
\label{tab-rq2-freq-categ}  
{\footnotesize
\begin{tabular}{p{1.55cm} p{1cm}  p{1cm} p{1cm}  p{1cm}  p{1cm}   p{1cm}  p{1cm} }
\hline
                      & \textbf{AGG}\  & \textbf{MINE}   & \textbf{STA}       & \textbf{EDU}    & \textbf{TRAK}  & \textbf{VOL}  & \textbf{EQU} \\  
\hline
\textbf{Bug categ.}     &                &                 &                    &                 &                &               &           \\  
\hline
\textbf{Algorithm}   & 6.8\%               & 6.7\%             & 22.4\%                 & 3.4\%              & 0.0\%          & 2.5\%              &  0.0\%  \\ 
\textbf{Data}   & 28.6\%            &  \cellcolor{green!50} 60.6\%           & 13.2\%                & 15.5\%       & 0.0\%              & 12.5\%             &  0.0\%  \\ 
\textbf{Dependency}  & 16.3\%             & 18.0\%            & 18.3\%                & 24.1\%               & 9.7\%          & 27.5\%            &  \cellcolor{green!50} 75.0\%  \\ 
\textbf{Document}  & 0.9\%              & 1.3\%               & 1.0\%                  & 0.0\%                  & 0.0\%         & 10.0\%            &  0.0\%  \\ 
\textbf{Performance}   & 2.7\%               & 2.0\%              & 0.0\%                 & 0.0\%                  & 3.2\%          & 0.0\%              &  0.0\%  \\ 
\textbf{Security}  & 1.8\%               & 0.0\%               & 3.0\%                 & 3.4\%                  & 6.4\%         & 12.5\%             &  0.0\% \\ 
\textbf{Syntax}  & 5.9\%              & 3.3\%               & 14.3\%                & 17.2\%                & 3.2\%           & 10.0\%              &  0.0\%   \\ 
\textbf{UI}      &  \cellcolor{green!50} 50.0\%            & 12.0\%             &  \cellcolor{green!50} 34.7\%               &  \cellcolor{green!50} 44.8\%                &  \cellcolor{green!50} 77.4\%        &  \cellcolor{green!50} 32.5\%             &    25.0\%  \\ 

\hline          
\end{tabular}
}
\end{table*}

\subsubsection{Rater Agreement and Verification} 
\label{res-rq2-rater} 

We report agreement rate for four steps: issue labeling, open coding, closed coding, and rater verification. 

\noindent \textit{Labeling issues as bugs}: While labeling collected issue reports as bug reports and non-bug reports the agreement rate is 96.5\% and the Cohen's Kappa is 0.9.    

\noindent \textit{Open coding to identify bug categories}: The first and second author respectively, identified 9 and 10 categories. The agreement rate is 72.7\%, and the Cohen's Kappa is 0.70, indicating `substantial' agreement~\citep{Landis:Koch:Kappa:Range}. The first author identified `database' as a category not identified by the second author. Upon discussion both authors agreed that `database' is related to data storage and belongs to the data category. The second author identified two additional categories `Public health data' and `Type errors'. After discussing the definitions of all categories both authors agreed that `Public health data' and `Type errors' can respectively, be merged with data and syntax.   

\noindent \textit{Closed coding to quantify bug category frequency}: During closed coding the first and second author mapped each project to an existing category. The agreement rate is 95.1\% and the Cohen's Kappa is 0.93. The authors disagreed on the labeling of 27 bug reports, which are resolved through discussion.    

\noindent \textit{Rater verification}: For the randomly selected 250 issue reports we allocate an additional rater who manually identified which of the issue reports are bug reports and non-bug reports. The Cohen's Kappa between the additional rater and the first author is 0.80, indicating `substantial' agreement~\citep{Landis:Koch:Kappa:Range}. The Cohen's Kappa between the additional rater and the second author is 0.84, indicating `perfect' agreement~\citep{Landis:Koch:Kappa:Range}. The agreement rate between the additional rater and the first and second author is respectively, 89.0\% and 93.0\%.        

We have also measured the agreement rate between an additional rater and the authors for categorizing bug reports. Cohen's Kappa between the additional rater and the first author for a randomly selected set of 250 bug reports is 0.65, indicating `substantial' agreement~\citep{Landis:Koch:Kappa:Range}. Cohen's Kappa between the additional rater and the second author for a randomly selected set of 250 bug reports is 0.68, indicating `substantial' agreement~\citep{Landis:Koch:Kappa:Range}. The agreement rate between the additional rater and the first and second author is respectively, 78.0\% and 81.6\%.             

\subsubsection{Resolution Time of Identified Bug Categories} 
\label{res-rq2-duration}

We provide bug resolution time as measured in hours for all bug categories in Table~\ref{tab-rq2-duration-all-bug-categ}. From Table~\ref{tab-rq2-duration-all-bug-categ} we observe that based on min and median bug resolution times security bugs take the longest to resolve, followed algorithm bugs. We also observe data bugs to take as long as 548 hours to resolve.  

A breakdown of bug resolution time across the six categories of projects is provided in Table~\ref{tab-rq2-duration-all-repo-categ}. The `All' row in Table~\ref{tab-rq2-duration-all-repo-categ} shows the minimum, median, and maximum bug resolution time for all bug categories measured in hours. 

In Table~\ref{tab-rq2-duration-all-repo-categ} we observe four instances where the minimum bug resolution time is less than 6 minutes ($< 0.1$ hours). One possible explanation can be practitioners' habit of opening a bug report after they have developed the fix for a bug~\citep{Wan:MSR2017, Thung:Lo:ISSRE2012}. In such cases, practitioners notice the bug early, construct the fix for the bug, and then submit the bug report by opening and closing the bug report promptly.

\begin{table}[]
\centering
\caption{Resolution time of identified bug categories. Resolution times is measured in hours. Median resolution time is highest for security bugs. }
\label{tab-rq2-duration-all-bug-categ}
{\footnotesize
\begin{tabular}{ p{4cm}  r r r}
\hline
\textbf{Bug category}   & \textbf{Min} & \textbf{Median} & \textbf{Max} \\
\hline
\textbf{Security}       & 1.240   & \cellcolor{green!50}  13.9  & 144.6  \\
\textbf{Algorithm}      & 0.041   & 13.5  & 172.7 \\
\textbf{Syntax}         & 0.004   & 12.1  & 174.2  \\
\textbf{UI}             & 0.003   & 11.8  & 254.2 \\
\textbf{Data}           & 0.003   & 8.4   & 548.0 \\
\textbf{Performance}    & 0.961   & 7.1    & 104.4  \\
\textbf{Dependency}     & 0.014   & 2.4    & 379.4\\
\textbf{Documentation}  & 0.013   & 1.4    & 76.8 \\
\hline
\end{tabular}
}
\end{table}

\begin{table}[]
\centering
\caption{Resolution time of bug categories grouped by project categories. We measure resolution time in hours. Median bug resolution time is highest for projects related to medical equipment software. }
\label{tab-rq2-duration-all-repo-categ}
{\footnotesize
\begin{tabular}{ p{7.0cm}  r r r}
\hline
\textbf{Project category}            & \textbf{Min} & \textbf{Median} & \textbf{Max} \\
\hline
\textbf{Medical Equipment}           & 5.0          & \cellcolor{green!50} 29.4            & 46.4 \\
\textbf{Volunteer Management System} & 0.013        & 21.1            & 174.2 \\
\textbf{User Tracking}               & 0.124        & 16.5            & 294.5  \\
\textbf{Education}                   & 0.121        & 11.2            & 294.5 \\
\textbf{Aggregation}                 & 0.003        & 8.7             & 379.4  \\
\textbf{Statistical Modeling}        & 0.004        & 7.2             & 168.3  \\
\textbf{Mining}                      & 0.005        & 2.5             & 548.1 \\
\hline
\textbf{All}                         & 0.003        & 7.4             & 548.0 \\ 
\hline 
\end{tabular}
}
\end{table}  

Median bug resolution duration for each project type and bug category is provided in Table~\ref{tab-rq2-duration-categ}. `AGG', `MINE', `STA', `EDU', `TRAK', `VOL' and `EQU' respectively, corresponds to the seven project categories: aggregation, mining, statistical modeling, education, user tracking, volunteer management system, and medical equipment. We observe median bug resolution time to vary across bug categories as well as for project categories. 

\begin{table*}[]
\captionsetup{justification=centering}
\caption{Median bug resolution time for each bug category and each project type measured in hours. `---' indicates categories for which no bug reports exist.  }  
\label{tab-rq2-duration-categ}  
{\footnotesize
\begin{tabular}{p{1.55cm} p{1cm}  p{1cm} p{1cm}  p{1cm}  p{1cm}   p{1cm}  p{1cm} }
\hline
                      & \textbf{AGG}\  & \textbf{MINE}   & \textbf{STA}       & \textbf{EDU}    & \textbf{TRAK}  & \textbf{VOL}  & \textbf{EQU} \\  
\hline
\textbf{Bug cat.}       &                &                 &                    &                 &                &               &           \\  
\hline
\textbf{Algorithm}   & 9.8                   & 10.8                  & 13.9                     & 10.1                   & ---                    & 13.5                &  ---  \\ 
\textbf{Data}   & \cellcolor{green!50} 12.2                 & 4.4                    & \cellcolor{green!50} 15.2                     & 17.0                   & ---                    & 42.0               &  ---  \\ 
\textbf{Dependency}  & 5.6                   & 0.1                    & 0.3                       & 4.5                    & 5.3                   & 2.9                  &  22.4  \\ 
\textbf{Document}  & 1.3                   & \cellcolor{green!50} 39.0                  & 1.5                       & ---                     & ---                   & 6.9                  &  ---  \\ 
\textbf{Performance}   & 7.1                   & 36.6                  & ---                       & ---                     & 1.5                   & ---                   &  ---  \\ 
\textbf{Security}  & 8.1                   & ---                     & 3.1                       & \cellcolor{green!50} 84.1                   & 13.9                 & 20.4                &  --- \\ 
\textbf{Syntax}  & 12.1                 & 4.7                     & 11.4                     & 8.6                     & 16.9                 & \cellcolor{green!50} 79.3                &  ---   \\ 
\textbf{UI}      & 8.3                  & 2.7                     & 13.1                     & 16.8                   & \cellcolor{green!50} 18.7                 & 21.9                 & \cellcolor{green!50} 46.4    \\ 

\hline          
\end{tabular}
}
\end{table*}


\section{Discussion} 
\label{discussion}

In this section, we first provide a summary of our findings in Section~\ref{disc-summary}. Next, we provide a discussion on the implications of our findings in Section~\ref{disc-implications}.

\subsection{Summary}
\label{disc-summary}

\begin{shaded} 


\hspace{-0.65cm} \textbf{\ul{Project category: Aggregation}} \\
\textit{Definition:} Aggregate COVID-19 data and present using visualizations  \\
\textit{Count :} 50 out of 129 (38.7\%) \\ 
\textit{Most frequent bug category:} UI bugs  \\
\textit{Median bug resolution time:} 8.7 hours  \\

\hspace{-0.65cm} \textbf{\ul{Project category: Mining}} \\
\textit{Definition:} Mine COVID-19 data   \\
\textit{Count :} 35 out of 129 (27.1\%) \\ 
\textit{Most frequent bug category:} Data bugs  \\
\textit{Median bug resolution time:} 2.5 hours  \\

\hspace{-0.65cm} \textbf{\ul{Project category: Statistical modeling}} \\
\textit{Definition:} Use of statistical models to make COVID-19 predictions    \\
\textit{Count :} 22 out of 129 (17.0\%) \\ 
\textit{Most frequent bug category:} UI bugs  \\
\textit{Median bug resolution time:} 7.2 hours  \\

\hspace{-0.65cm} \textbf{\ul{Project category: Education}} \\
\textit{Definition:} Educate people about COVID-19   \\
\textit{Count :} 9 out of 129 (6.9\%) \\ 
\textit{Most frequent bug category:} UI bugs  \\
\textit{Median bug resolution time:} 11.2 hours  \\

\hspace{-0.65cm} \textbf{\ul{Project category: User tracking}} \\
\textit{Definition:} Track user data related to COVID-19   \\
\textit{Count :} 9 out of 129 (6.9\%) \\ 
\textit{Most frequent bug category:} UI bugs  \\
\textit{Median bug resolution time:} 16.5 hours  \\

\hspace{-0.65cm} \textbf{\ul{Project category: Volunteer management}} \\
\textit{Definition:} Efficiently manage volunteering effort related to COVID-19   \\
\textit{Count :} 7 out of 129 (5.4\%) \\ 
\textit{Most frequent bug category:} UI bugs  \\
\textit{Median bug resolution time:} 21.1 hours  \\

\hspace{-0.65cm} \textbf{\ul{Project category: Medical equipment}} \\
\textit{Definition:} Source code for design and implementation of medical devices   \\
\textit{Count :} 3 out of 129 (2.3\%) \\ 
\textit{Most frequent bug category:} Dependency bugs  \\
\textit{Median bug resolution time:} 29.4 hours  \\


\end{shaded}

\subsection{Implications} 
\label{disc-implications} 

We discuss the implications of our findings below: 

\textbf{Security and privacy implications of user tracking software}: From Table~\ref{tab-res-rq1-categ-atts} we observe 9 projects to be related with user tracking. While the benefits of user tracking software has been documented for countries, such as Russia and South Korea~\citep{covid19:privacy:violation}, this category of software can have negative impacts on privacy of end-users. Data generated from user tracking software can be leveraged for marketing purposes. To preserve privacy of user data in user tracking software we make the following recommendations: 

\begin{itemize}
    \item{Policy makers should construct policies specific to COVID-19 software that collects user data.}
    \item{Practitioners who develop user tracking software should leverage existing privacy policy frameworks, such as the `National Institute of Standards and Technology (NIST) Privacy Framework'~\citep{nist:privacy:framework}.}
    \item{Privacy researchers can build tools that will automatically detect and report privacy policy violations.}
\end{itemize}

Evidence from Table~\ref{tab-rq2-freq-categ} shows that security bugs to exist for user tracking software. We advocate security researchers to systematically investigate if user tracking software includes security bugs. Recent news articles suggest that user tracking software, such as contract tracing apps may become more and more prevalent as Apple and Google are already providing frameworks to build software that tracks user data.~\citep{apple:tracing:api}. Our hypothesis is that availability of these frameworks will facilitate rapid development and deployment of mobile apps that collect user data. Security weaknesses in these apps can provide malicious users opportunity to conduct large-scale data breaches. We notice anecdotal evidence in this regard: a researcher has identified vulnerabilities in a user tracking app that could leak user location data~\citep{covid19:tracing:india}. Panelists at EuroCrypt 2020, a cryptography research conference, discussed limitations of user tracking mobile apps for COVID-19 with respect to API design, indoor location tracking, and informing users about privacy risks~\citep{eurocrypt2020:panel:contract:tracing}~\citep{eurocrypt2020:videos:contract:tracing}.

\textbf{Towards constructing correct statistical models}: From Section~\ref{res-rq2-categ} we have observed statistical modeling bugs to exist. Bugs related to statistical modeling can be consequential because based on the predictions generated by statistical models, policy makers enforce public health policies. One possible explanation for buggy statistical models can be attributed to the quality of datasets using which statistical models are build~\citep{covid19:modeling}. For example, fatality prediction models that are built using the `Diamond Princess Cruise Ship Dataset' may not be applicable for a specific geographic region with low population density. Another possible explanation can be a lack of context and knowledge related to public health specific that hinders model builders to identify appropriate independent variables to construct the models. Incorrect estimation of hospital beds from our discussion in Section~\ref{res-rq2-categ} is one example. Other examples of independent variables related to public health includes staff availability, count of known cases, hospitalization rate etc.~\citep{covid19:hospitilization}. According to a health expert~\citep{covid19:hospitilization}, statistical models that predicted 2.4 million US residents to die, assumed a hospitalization rate of 15-20\%, which in reality was 5\%.

Based on our findings and above-mentioned explanations we make two recommendations: 

\begin{itemize}
\item{\textit{Automated testing for COVID-19 modeling}: We hope to see novel research in the domain of COVID-19 that will test the correctness of constructed statistical models used in forecasting in an automated manner. In recent years, we have seen research efforts that test deep learning models~\citep{deeptest:baishakhi, sosp:test:deep, issre:test:deep}. We expect similar research pursuits for COVID-19 statistical modeling.   
}

\item{\textit{Better synergies between data science and public health practitioners}: Construction and verification of COVID-19 statistical modeling should involve practitioners from public health and data science. Public health practitioners within a specific locality can provide necessary context that data scientists can incorporate in their statistical models.    

}
\end{itemize}   

\textbf{Implications for Educators}: Our findings have implications for educators involved in teaching the following topics: 

\begin{itemize}
\item{\textit{Data science}: Educators who teach data science can use the examples of statistical modeling bugs to highlight the value of considering the full context and related limitations that accompany statistical modeling.   
}

\item{\textit{Information security and privacy}: User tracking software can be discussed in information security and privacy courses to demonstrate the value of protecting user data. Such discussion can also include privacy policy frameworks that are already in place, such as the NIST Privacy framework~\citep{nist:privacy:framework}.  
}

\item{\textit{Software engineering}: Our categorization of bugs related to COVID-19 software development can be discussed to demonstrate that understanding and repair of bugs requires contextualization.   
}
\end{itemize}

\textbf{Benchmark for practitioners and researchers}: Tables~\ref{tab-rq2-freq-all}---~\ref{tab-rq2-duration-categ} can be used  as a measuring stick by practitioners and researchers who are involved with COVID-19 software projects. Practitioners can estimate their bug resolution efforts by comparing median resolution times for bugs in their COVID-19 software projects to that of Tables~\ref{tab-rq2-duration-all-bug-categ},~\ref{tab-rq2-duration-all-repo-categ}, and~\ref{tab-rq2-duration-categ}. 

Compared to prior work related to blockchain and machine learning~\citep{Thung:Lo:ISSRE2012, Wan:MSR2017}, median bug resolution time is lower for COVID-19 software projects. We provide two possible explanations: one possible explanation can be related to the sense of urgency. Practitioners may have realized that bugs in COVID-19 software projects could hamper the analysis or mitigation of COVID-19, and therefore, needs immediate attention. Another possible explanation can be the limitations of our dataset. The age of our software projects do not exceed four months and that may have biased median bug resolution time. We advocate for future research that will confirm or refute our explanations.  

\textbf{Recurrence-related implications}: In Section~\ref{bg}, we have discussed evidence related to recurrence of COVID-19. We hypothesize that COVID-19's recurrence will lead to more COVID-19 software building. Whether or not our findings hold for these newly constructed COVID-19 software can be validated through a replication of our paper. We expect to observe more categories of COVID-19 software projects as well as more bug categories that could occur with varying distributions.

\textbf{Mitigation strategies}\label{disc-mitgation}: We suggest the following mitigation strategies for each of the identified bug categories: 

\begin{itemize}[leftmargin=*]
    \item{\textit{Algorithm}: Algorithm bugs can be mitigated through peer reviews. One sub-category of algorithm bugs is bugs related to statistical models built for COVID-19. Mitigation of bugs in statistical models may require synergies between practitioners from data science and public health.  
    }
    \item{\textit{Data}: Data bugs can be mitigated by adequate accumulation of data, a strategy which requires non-trivial amount of effort. We advocate for community effort so that COVID-19 data sources are curated. 
    }
    \item{\textit{Dependency}: Practitioners can mitigate dependency bugs by using automated tools, such as Dependabot~\footnote{https://dependabot.com/}.  
    }
    \item{\textit{Documentation}: Practitioners can mitigate documentation bugs through peer review and documentation-specific linters that can alert practitioners when incorrect information is specified in README files. 
    }
    \item{\textit{Performance}: Researchers can use research tools, such as Toddler~\citep{darko:perf:toddler} and Clarity~\citep{clarity:perf:bugs} to detect performance bugs in software source code. Furthermore, performance bugs in UIs can be detected by record and replay tools, such as Pounder as suggested by researchers~\citep{gui:perf:replay}.  
    }
    \item{\textit{Security}: We advocate practitioners to use automated security scanning tools applicable for the language used in the COVID-19 software repository. Prevalence of security bugs can be detected and mitigated by applying security scanning tools.  
    }
    \item{\textit{Syntax}: We advocate practitioners to use existing static analysis tools to mitigate syntax bugs.  
    }
    \item{\textit{UI}: Practitioners can mitigate UI bugs by using automated UI testing tools, such as Selenium~\footnote{https://www.selenium.dev/}. We have noticed UI bugs related accessibility issues, which can be mitigated by automated tools, such as Applause~\footnote{https://www.applause.com/}.  
    }
\end{itemize}


\section{Threats to Validity} 
\label{threats}

We describe the limitations of our paper as following: 

\textbf{Conclusion validity}: We have used raters who derived the software and bug categories. Both raters are authors of the paper. Our derived categoires are susceptible to the authors' bias. We mitigate this limitation by allocating another rater who is not the author of the paper who verified our ratings.   

Our categories might not be comprehensive because our categorization for projects and bugs is limited to the dataset that we collected. The bug resolution time could be limiting as our dataset includes projects that have a duration of four months.

We use the topic `covid-19' to identify and filter COVID-19 software projects from GitHub. Any software project that is not labeled as `covid-19' will not be included in our dataset.  

Our datasets have limited lifetime as the COVID-19 was discovered in December 2019, and the lack of maturity in our datasets may influence our analysis. We mitigate this limitation by identifying projects using a filtering criteria so that we can identify projects with sufficient development activity.  

\textbf{Internal validity}: For RQ1 and RQ2 we use ourselves, the authors of the paper, as raters who conduct open and closed coding on README files and bug reports. Our research is susceptible to mono-method bias, as our categorization and labeling may be influenced by the authors' implicit expectations and hypotheses about the study.

\textbf{External validity}: Our findings are not comprehensive. We have not analyzed projects hosted outside GitHub and private projects hosted on GitHub. We mitigate this limitation by analyzing 129 software projects that belong to 7 categories.


\section{Conclusion} 
\label{conclusion} 

The COVID-19 pandemic has impacted people all over the world causing thousands of deaths. Software practitioners have joined the fight in combating the spread and mitigating the dire consequences of COVID-19. An understanding of COVID-19 software categories and software bugs can give us clues on how the software engineering community can help even further in combating COVID-19.

We conduct an empirical study with 129 COVID-19 software projects hosted on GitHub. We identify 7 categories of software projects: aggregation, mining, statistical models, education, volunteer management, user tracking, and medical equipment. By applying open coding on 550 bug reports, we identify 8 categories of bugs: algorithm, data, dependency, documentation, performance, security, syntax, and UI. We observe bug category frequency to vary with project categories, e.g., for mining projects data-related bugs is the most frequently occurring category.    

Our findings have implications for educators, practitioners, and researchers. Educators can use our categorization of COVID software projects and related bugs to educate students about the security and privacy implications of COVID-19 software. Privacy researchers can build tools that will check if user tracking software related to COVID-19 are not leaking user data. Practitioners in the data science domain can learn from our categorization of statistical modeling bugs to understand limitations of constructed statistical models and verify underlying assumptions that accompany constructed statistical models. Based on our findings we also advocate for better synergies between data scientists and public health experts so that statistical modeling bugs can be mitigated. We hope our paper will advance further research in the domain of COVID-19 software.  


\begin{acknowledgements}
We thank the PASER research group members at Tennessee Technological University for their useful feedback. We also thank Farzana Ahamed Bhuiyan of Tennessee Technological University for her help as an additional rater. 
\end{acknowledgements}

\balance 
\bibliographystyle{spbasic}  
\bibliography{covid19}

\end{document}